\documentclass[aps,twocolumn]{revtex4}

\newcommand{\up}{\uparrow}
\newcommand{\down}{\downarrow}
\newcommand{\de}{\partial}
\input epsf.tex
\begin{document}
\title{Diluted magnetic semiconductor quantum dots: an extreme 
sensitivity of the hole Zeeman splitting on the aspect ratio of the
confining potential}
\author{F.V.Kyrychenko\footnote{Corresponding author: F.V.Kyrychenko,
Institute of Physics Polish Academy of Sciences, Al. Lotnik\'ow 32/46,
02-668 Warsaw, Poland.
Tel.: +48-22-843-1331; fax: +48-22-843-0926;
e-mail: kirich@ifpan.edu.pl}, J.Kossut}
\address{Institute of Physics, Polish Academy of Sciences,
Al. Lotnik\'ow 32/46, 02-668 Warsaw, Poland}

\begin{abstract}
The valence band states confined in 
infinitely deep quantum dots made of diluted magnetic
semiconductors (DMS) are considered theoretically. A complex
anisotropic structure of the valence bands in DMSs with cubic symmetry
described by the full
Luttinger Hamiltonian is taken into account. It is found
that the Zeeman splitting is very sensitive to the shape of
the confining potential and, in particular, to its orientation relative
to the direction of an external magnetic field.
This sensitivity has its origin in a mixing of different spin components
of a hole wave function which takes place for finite hole wave
vectors ${\bf k}$. Several consequences of the effect are discussed,
including a possibility to control the inter-dot tunneling
by an external
magnetic field. It is shown also that the polarizations of  
optical transitions
in a single DMS quantum dot depend on details of geometry of its 
confining potential as well as on the strength of the magnetic field.
\end{abstract}


\maketitle

\section{Introduction}
From the moment of initial stages of their investigation
in mid-seventies (see, e.g.,
\cite{ryabch76,bastard77,jk77}) until present diluted magnetic
semiconductors (DMSs) and quantum structures involving them, continue
to attract a considerable attention of the
scientific community (see, e.g., \cite{SSv25,dietl}). 
The interest has recently gained impetus with new focus on
spintronic materials (see, e.g., recent review \cite{oestreich}). 
The part of this interest stems from the fact that the
presence of an exchange interaction between the band carriers and
electrons from the partially-filled $d$- or $f$-shells 
of magnetic ions in DMSs gives rise to a unique possibility 
to control electrical and optical
properties of quantum structures by means of an external magnetic field
in a range difficult to achieve in other materials.

This $s$-$d$ exchange interaction, in the mean field approximation,
can be cast \cite{SSv25} into Zeeman-like form
\begin{equation}
\label{Hsd}
\hat{H}_{s-d}=N_0\alpha x \langle {\bf \hat{S}} \rangle\cdot {\bf \hat{s}},
\end{equation}
in the case of the conduction electrons, and the $p$-$d$ interaction in
the same approximation
\begin{equation}
\label{Hpd}
\hat{H}_{p-d}=\frac{1}{3}N_0\beta x \langle {\bf \hat{S}} \rangle
\cdot {\bf \hat{J}},
\end{equation}
in the case of the valence band holes. Here ${\bf \hat{s}}$ and 
${\bf \hat{J}}$ are operators of the electron spin 
and the total angular momentum of the hole, respectively, $x$ is
the molar fraction of magnetic ions introduced substitutionally
into the semiconductor matrix and $\langle {\bf \hat{S}} \rangle$
represents their average localized spin.

The corresponding exchange constants $N_0 \alpha$
for the conduction band states and $N_0 \beta$ for the valence 
band states in majority
of II-VI DMSs are usually of opposite signs \cite{blinkac}. 
Usually also, the absolute value of $N_0 \beta$ is considerably 
greater than $N_0 \alpha$. For example, in
$\rm Cd_{1-x}Mn_x Te$ $N_0 \alpha=0.22$ eV and
$N_0 \beta=-0.88$ eV \cite{gajplanelfishman}. Thus the giant spin
splitting (GSS) of excitonic states in DMSs is mainly determined by the
spin splitting of the {\em hole states}.

The excitonic GSS in DMS quantum structures is determined by the
hole splitting to an even greater extent than in bulk crystals.
The reason for this is that the
electron and the hole states in low-dimensional semiconductor structures
are energy dependent mixtures (stemming from the 
$k$-dependence of the mixing in
the bulk) of the Bloch amplitudes with $\Gamma_6$ and
$\Gamma_8$ symmetries of the center of the Brillouin zone point 
group representations.
Since the negative exchange constant $N_0\beta$ is several times
greater in magnitude than the positive $N_0\alpha$, even a small
admixture of $\Gamma_8$ states to $\Gamma_6$ state can reduce
significantly (up to 30\% in the case of
quantum wells) the value of total effective exchange constant
of the conduction electrons \cite{merkulov}. For quantum wires (QWRs) and,
especially, for quantum dots (QDs) this reduction of the
conduction band exchange constant is expected to be even more significant.
A degree of a reciprocal process of a reduction of 
the value of the effective
exchange constant of the holes due to an admixture of $\Gamma_6$ 
symmetry to $\Gamma_8$ states is less pronounced
because of the mentioned difference of the magnitudes of the 
exchange constants $N_0\alpha$ and $N_0\beta$.
Therefore, the exciton spin splitting in
DMS quantum structures, which combines the splitting of the electron
and the hole comprising it, is dominated by what happens 
in the valence band.

An example of such processes, that affects the
Zeeman splitting of the holes, is an additional heavy 
hole-light hole mixing induced by the QD confining potential. 
This has been noted and exploited in Ref.~\cite{bhat_hhlh} where 
author used an exactly
solvable model, namely of the Luttinger Hamiltonian in a 
spherical approximation with spatial confinement given by a spherically 
symmetric quantum dot potential. The reduction of the Zeeman splitting
of an optically active ground state of an exciton due to heavy-light hole
mixing was found to be independent of the QD size. The degree of
the reduction depends only
on the relation between the heavy and the light hole effective masses in
the crystal and for CdTe it amounts to about $\rho\approx 0.8$.

In the present paper we pursue the study of the consequences of 
the heavy hole-light hole
mixing induced by the localization of the holes by a QD potential.
However, we concentrate mainly on the dependence of the mixing on the
{\em shape} or, more specifically, on the aspect ratio 
of the confining potential. We demonstrate that in different
QDs, giving rise to similar confinement energies, the symmetry and,
thus, the total angular momentum of the hole ground
state might be completely different. To illustrate this point, let 
us imagine that a characteristic
length of a quantum dot in, say, $z$-direction $L_z$ is much 
smaller than that in
$x$- and $y$-directions $L_z\ll L_x,L_y$. The situation becomes then
similar to the case of a quantum well (QW), where the ground state 
of the hole is purely of a heavy
hole in character with $J_z=\pm 3/2$. In the opposite case, when
$L_z\gg L_x,L_y$ the structure resembles
a QWR with the ground state of holes being mainly the light hole in
character due to the greater masses in $x$-$y$ plane \cite{sercel,PRB}. 
Yet, from the point
of view of the confinement energy only both these two extreme cases
may be the same.

One of the consequences of different symmetries of the hole ground states
in different structures having the same confinement energy is a 
strong dependence
of the expectation value $\langle \hat{J}_z \rangle$ on the shape
of the confining potential. On the other hand, the Zeeman splitting of
the hole states produced by the interaction Hamiltonian (\ref{Hpd})
in the first order of perturbation theory is proportional
to the value of $\langle \hat{J}_z \rangle$ (we shall always assume that
external magnetic field ${\bf B}$ that induces the magnetization 
appearing in Eqs.~(\ref{Hsd}) and (\ref{Hpd}) is along the $z$-axis).
One can expect, thus, in the case of DMSs a strong dependence of the 
effective exciton spin splitting on the aspect ratio of 
quantum dots. We will show also that due to the strong anisotropy 
of the hole states the very orientation of the QD with respect to
an external magnetic field affects dramatically the Zeeman splitting.

It was shown in Ref.~\cite{bhat_k} that due to
the $k$-dependence of $s(p)$-$d$ exchange constants their values in
quantum structures are reduced to a certain extent. Nevertheless,
for the sake of simplicity , to describe $s$-$d$ and $p$-$d$ 
exchange interactions we use in this paper the values of these constants 
equal to their values in the bulk. This simplification becomes invalid
only in very small QDs (with radius smaller than about 30$\AA$), while
here we focus our attention on much greater structures and we show that
even in those the dependence of the Zeeman splitting on
their geometry is significant. Admittedly, for small QDs 
one should replace bulk
$N_0\beta$ by the values of $p$-$d$ exchange constants, calculated
in \cite{bhat_k}. This does not change, of course, 
the qualitative results obtained in the present paper.

To sum up, we are considering here the shape-induced changes of
$\langle \hat{J}_z \rangle$, and their various consequences, 
in DMS QDs while 
neglecting the variation of exchange constant $N_0 \beta$  itself
since we expect it to be of smaller importance.

\section{Theoretical model}

We consider the valence band states in a diluted magnetic semiconductor
QD in a shape of a rectangular parallelepiped with 
dimensions $L_x$, $L_y$, 
and $L_z$. This particular choice of the form of confining potential
is motivated by two reasons. First, by changing the relations 
between $L_x$, $L_y$ and $L_z$ one can control the
asymmetry of the confining potential. This is convenient both for 
calculations and for interpretation of the results obtained.
The second reason is that the symmetry of QD potential
in this particular case is rather low. In a general case, 
this is $D_{2h}$ 
symmetry. Such a low symmetry prevents appearance 
of additional constants of motion, such as the angular momentum
which "creeps in" in the case of spherically symmetric QDs.

For simplicity we consider the case of a dot with infinitely high
potential barriers surrounding it. The parameters $L_x$, $L_y$ 
and $L_z$ can be treated then
as characteristic length scales of the hole localization in a real QD 
(i.e., with finite potential barriers) in $x$, $y$ and $z$-directions.

Our basic starting Hamiltonian is the Luttinger 
Hamiltonian \cite{luttinger} describing the hole states
in a cubic semiconductors which can be presented in the form
\begin{eqnarray}
\label{HamLut}
\hat{H}_L &=& \frac{\hbar^2}{2m_0}
\left[\left(\gamma_1+\frac{5}{2}\gamma_2
\right)\left(\hat{K}_x^2+\hat{K}_y^2+\hat{K}_z^2\right)\right. \nonumber \\
& - & 2\gamma_2
\left(\hat{K}_x^2 J_x^2+\hat{K}_y^2 J_y^2+\hat{K}_z^2 J_z^2\right) \\
& - & 2\gamma_3\biggl(\hat{K}_x\hat{K}_y(J_xJ_y+J_yJ_x)+\hat{K}_x\hat{K}_z
(J_xJ_z+J_zJ_x) \nonumber \\
& + &  \hat{K}_z\hat{K}_y(J_zJ_y+J_yJ_z)\biggr)\biggr], \nonumber
\end{eqnarray}
where $\hat{K}_x=-i\de/\de x$, $\hat{K}_y=-i\de/\de y$,
$\hat{K}_z=-i\de/\de z$, with the $z$-axis is parallel to
$[001]$ crystallographic direction, $\gamma_i$ are the Luttinger
parameters, while $J_x$, $J_y$ and $J_z$ being
$4\times 4$ operator matrices of the projections of the angular 
momentum $J=3/2$ in the
basis of $\Gamma_8$ Bloch amplitudes
\begin{eqnarray}
\psi_{3/2} & = & \frac{1}{\sqrt{2}}(X+iY)\up, \label{+3/2} \\
\psi_{1/2} & = &
\frac{1}{\sqrt{6}}\biggl[(X+iY)\down -2Z\up\biggr], \label{+1/2} \\
\psi_{-1/2} & = &
\frac{1}{\sqrt{6}}\biggl[-(X-iY)\up -2Z\down\biggr], \label{-1/2} \\
\psi_{-3/2} & = & -\frac{1}{\sqrt{2}}(X-iY)\down. \label{-3/2}
\end{eqnarray}

We expand the orbital part of each of the spin components
(\ref{+3/2})-(\ref{-3/2}) of the hole wave function in the
basis of the solutions of the Schr\"{o}dinger equation for an 
infinitely deep QD.  The hole wave function is then represented by
\begin{eqnarray}
\label{basis}
\Psi&=&\sum_{J_z=-3/2}^{+3/2} \sum_{n=1}^{N_x}\sum_{m=1}^{N_y}
\sum_{l=1}^{N_z} A^{n,m,l}_{J_z}\sqrt{\frac{8}{L_x L_y L_z}}\nonumber \\
&\times& \sin(n k_x x)\sin(m k_y y) \sin(l k_z z)\cdot\psi_{J_z},
\end{eqnarray}
where $k_x=\pi/L_x$, $k_y=\pi/L_y$, $k_z=\pi/L_z$,
$N_x$, $N_y$ and $N_z$ are (finite) numbers of the basis functions 
used in the expansion and
$A_{J_z}^{n,m,l}$ being numerical coefficients. Let us note that 
in general $\Psi$ is not an eigenfunction of $J_z$ or any other
component of ${\bf J}$.

For the case of the hole ground state let us first limit the 
expansion (\ref{basis}) 
to the terms with $m=n=l=1$. In this case the orbital part of the hole
envelope has the form
\begin{equation}
\label{orbit}
\Phi(x,y,z)=\sqrt{\frac{8}{L_x L_y L_z}}
\sin(k_x x)\sin(k_y y) \sin(k_z z).
\end{equation}
This approximation is equivalent to neglecting of term proportional
to $\gamma_3$ in the Hamiltonian (\ref{HamLut}) or, more strictly, to 
a reduction of the spherical term $\propto\gamma_3({\bf\hat{K}}
\cdot{\bf J})^2$ to a one with cubic symmetry $\propto\gamma_3(\hat{K}_x^2 
J_x^2+\hat{K}_y^2 J_y^2+\hat{K}_z^2 J_z^2)$. In fact, the 
terms proportional
to $\hat{K}_{\alpha}^2$  in Eq.(\ref{HamLut}) do not mix states with
different orbital functions in the expansion (\ref{basis}) and 
the hole ground state has the form (\ref{orbit}). 

With the above approximation the last term in Eq.(\ref{HamLut}), 
being linear in the operators
$\hat{K}_{\alpha}$, has only vanishing matrix elements.
At the same time, the second term in Eq.(\ref{HamLut}) is
proportional to the squares of the matrices $J_{\alpha}$. Thus, the
non-vanishing matrix elements of this term involve states 
with $\Delta J_z=\pm 2$ only. As a result, the $4 \times 4$ 
matrix of the
Hamiltonian (\ref{HamLut}) splits into two $2 \times 2$ identical
submatrices in the basis $(+3/2,-1/2)$ and $(-3/2,+1/2)$. These matrices
have the form
\begin{equation}
\label{2x2}
H=\frac{\hbar^2}{2 m_0}
\left(
    \begin{array}{cc}
      P+Q & R \\ R & P-Q
    \end{array}
\right),
\end{equation}
where
\begin{eqnarray*}
&& P=\gamma_1 (k_x^2+k_y^2+k_z^2) \\
&& Q=\gamma_2 (k_x^2+k_y^2-2 k_z^2) \\
&& R=-\sqrt{3}\gamma_2 (k_x^2-k_y^2).
\end{eqnarray*}

In the case of a strictly cubic QD ($k_x$=$k_y$=$k_z$), 
the ground state of the hole is four-fold degenerate. The hole
states can be characterized by the projection of the angular momentum 
$J_z=\pm 3/2,\,\pm 1/2$.
We can remove this degeneracy by reducing the symmetry of the
Hamiltonian from $O_h$ to $D_{4h}$ ($k_z \neq k_x = k_y$).
This leads to the heavy hole-light hole splitting. $J_z$
remains a good quantum number. The hole ground state doublet is then
the heavy hole doublet with $J_z=\pm 3/2$ if $k_x^2+k_y^2<2 k_z^2$,
while it corresponds to the light hole doublet with $J_z=\mp 1/2$, if 
$k_x^2+k_y^2>2k_z^2$.

In the general case, $k_x\neq k_y \neq k_z$, (the symmetry of the
Hamiltonian being
$D_{2h}$) the hole states are the mixtures of the components
with $J_z=\pm 3/2$
and $J_z=\mp 1/2$. In the absence of an external magnetic 
field the ground state with the energy
\begin{equation}
E=\frac{\hbar^2}{2 m_0} \left( P-\sqrt{Q^2+R^2} \right)
\end{equation}
is twofold degenerate. This effective spin doublet is characterized by
an expectation value of the projection of the total angular momentum
\begin{equation}
\label{<Jz>}
\langle J_z \rangle = \pm\left( \frac{2}{1+\left(\delta+
\sqrt{\delta^2+1}\right)^2}-\frac{1}{2}\right),
\end{equation}
where
\begin{equation}
\delta=\frac{Q}{|R|}=\frac{k_x^2+k_y^2-2 k_z^2}{\sqrt{3}|k_x^2-k_y^2|}.
\end{equation}
For degenerate states the choice of the wave functions is not unique.
To obtain Eq.(\ref{<Jz>}) we use the correct zero-order
functions with respect to the perturbation $G \hat{J}_z$.

For each value of
$k_x$ and $k_y$ it is possible to find a value of $k_z$ that leads to
vanishing of $\langle J_z \rangle$. The corresponding condition
is $\delta = 1/\sqrt{3}$. In the case of $k_x > k_y$
to satisfy this equality one has to choose $k_z=k_y$. This result easy
to see: the equality $k_z=k_y$ under the condition
$k_x>k_y$ means that the hole ground state is a superposition of the
states with $J_x=\pm 3/2$  only. The average value $\langle J_z \rangle$
for such states is equal to zero, similarly to the case when
the states are the mixtures of states with $J_z=\pm 3/2$ and then 
$\langle J_x \rangle$ and $\langle J_y \rangle$ vanish.

It is seen, thus, that the expectation value of $J_z$
for the hole ground state in a QD varies in a wide range from
$\langle J_z \rangle=\pm 3/2$ to $\langle J_z \rangle=0$ depending on
the exact shape of the confining potential, i.e., relation between its
extension in three crystallographic directions.

The situation changes somewhat if we take into account
more than one term in the expansion (\ref{basis}). The projection $J_z$ 
ceases to be a good quantum number for the holes even in the case when
$k_x=k_y$. This is due to the presence of the term 
proportional to $\gamma_3$ in Eq.(\ref{HamLut}). 
Moreover, the hole states now are the mixtures of
all four spin components.
Nevertheless, as before, one can 
vary effectively the value of $\langle J_z \rangle$ in the ground
hole state from nearly $\pm 3/2$ to $0$ by changing the shape of the
confining potential. 
\begin{figure}
\epsfysize=6cm \epsfbox{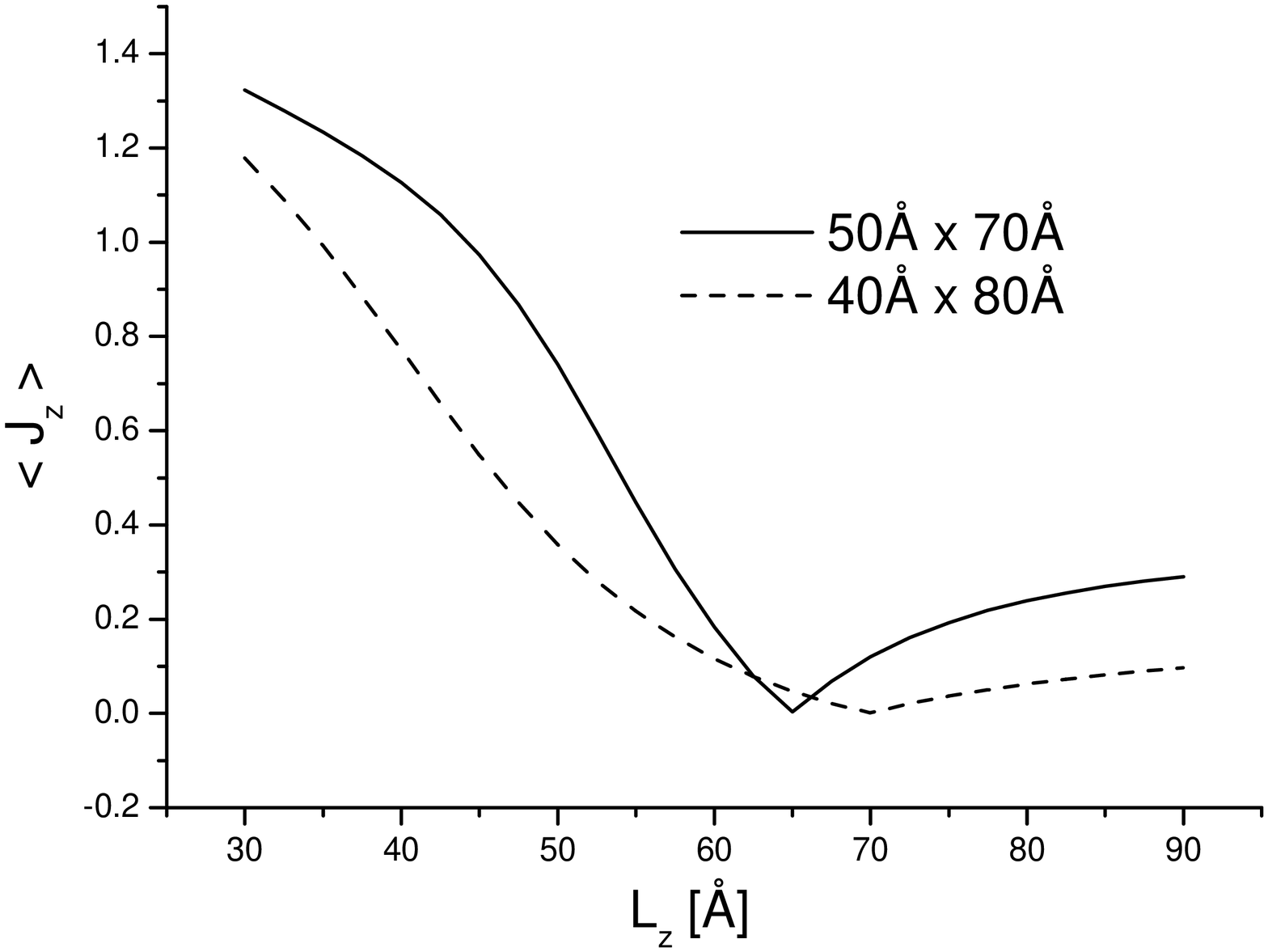}
\caption{\label{Jz} The expectation value of the projection of the total
angular momentum $\langle \hat{J}_z \rangle$ as a function of 
the QD size in the $z$-direction. The solid line corresponds to
the structure with $L_x=50\AA$ and $L_y=70\AA$, while the
dashed one corresponds to the structure with $L_x=40\AA$ and $L_y=80\AA$.}
\end{figure}
To illustrate this feature
we plot in Fig.~\ref{Jz} the
expectation value of the projection of the total angular 
momentum $J_z$ for the
ground spin doublet in the QD having $L_x=50\AA$ and
$L_y=70\AA$ (solid line) and $L_x=40\AA$ and
$L_y=80\AA$ (dashed line) as a function of $L_z$. Hereafter we use the
number of basis functions in the expansion (\ref{basis}) $N_x=N_y=N_z=6$,
which were checked to be sufficient to ensure numerical accuracy (about
1 meV in calculation of hole energy). In our calculations we use 
the CdTe Luttinger parameters $\gamma_1=5.3$, 
$\gamma_2=1.62$ and $\gamma_3=2.1$ \cite{gamma}. 
Figure~\ref{Jz} shows that the value $\langle J_z \rangle$ is, 
in fact, very sensitive to the form of the quantum dot potential.

\begin{figure}[h]
\epsfysize=6cm \epsfbox{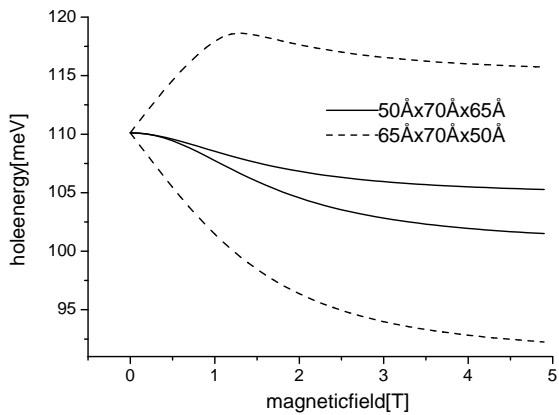}
\caption{\label{zeem} The Zeeman splitting of the hole ground states in
$\rm Cd_{0.97}Mn_{0.03}Te$ DMS quantum dots in the presence of a
magnetic field
${\bf B}\parallel z$ at $T=2K$. The solid lines correspond to the 
structure with
$L_x=50\AA$, $L_y=70\AA$ and $L_z=65\AA$, the dashed 
lines -- to the structure with
$L_x=65\AA$, $L_y=70\AA$ and $L_z=50\AA$.}
\end{figure}

As mentioned above, the value of the Zeeman splitting of the
valence band states
in the magnetic field ${\bf B}\parallel z$ is proportional, 
in the first order of perturbation, to $\langle J_z \rangle$.
However, the perturbation theory is not quite justified if the value of
the Zeeman splitting becomes comparable to the energy separation 
between the size-quantized states in a QD. To
calculate the Zeeman splitting in such case we perform a numerical
diagonalization of the full Hamiltonian matrix for the holes, which 
is represented by the sum of the
Luttinger Hamiltonian (\ref{HamLut}) and the $p$-$d$ exchange interaction
(\ref{Hpd}). Note, that we neglect here the direct influence of the
magnetic field on the hole states. The magnetic field in our model 
just induces the magnetization on the sample, which is proportional,
in turn, to
the average spin $\langle {\bf \hat{S}} \rangle$ of the magnetic ions
appearing in
(\ref{Hpd}). In the calculations that are presented below 
we use the parameterization of the value
$\langle {\bf \hat{S}} \rangle$ in real DMS structures, given 
in \cite{gaj_profile}.

\section{Discussion}

The evolution of the positions of the two lowest energy levels in
the valence band in $\rm Cd_{0.97}Mn_{0.03}Te$ DMS quantum
dots having $L_x=50\AA$,
$L_y=70\AA$ and $L_z=65\AA$ (solid lines) and $L_x=65\AA$,
$L_y=70\AA$ and $L_z=50\AA$ (dashed lines) in the magnetic field
${\bf B}\parallel z$ is presented in Fig.~\ref{zeem}. 
Such QDs provide exciton confinement energy of about 400 meV, that is of
the correct magnitude as compared to existing experimental 
observations in real self-assembled QDs made of this material
\cite{QD_energy}. In a 
small magnetic field the value of the Zeeman splitting of the ground 
state is proportional to the expectation
value of the projection of the total angular momentum of
the hole on the direction
of the magnetic field. This quantity is
$\langle \hat{J}_z \rangle=0.003$ for
the first structure and $\langle \hat{J}_z \rangle=1.2$ for the second
structure, which is just the first one merely rotated by $\pi /2$ around
the $y$-axis. Due to the fact that $J_z$ is not a good quantum number,
the magnetic field ${\bf B}\parallel z$ induces the mixing of
the hole states corresponding to different size-quantized levels.
With an increase of the
magnetic field the admixture of the higher energy states to the ground
levels becomes significant and it produces a deviation of the value
of the Zeeman splitting from that predicted by the first order of
perturbation theory. Nevertheless, Fig.~\ref{zeem} shows
that due to a strong dependence
of $\langle \hat{J}_z \rangle$ on the actual geometry
it is possible to obtain DMS quantum dots with the same,
homogeneous distribution of the magnetic impurities, that have the same
space-quantized energies but, nevertheless, differ essentially in 
their effective g-factors.

\begin{figure}[h]
\epsfysize=6cm \epsfbox{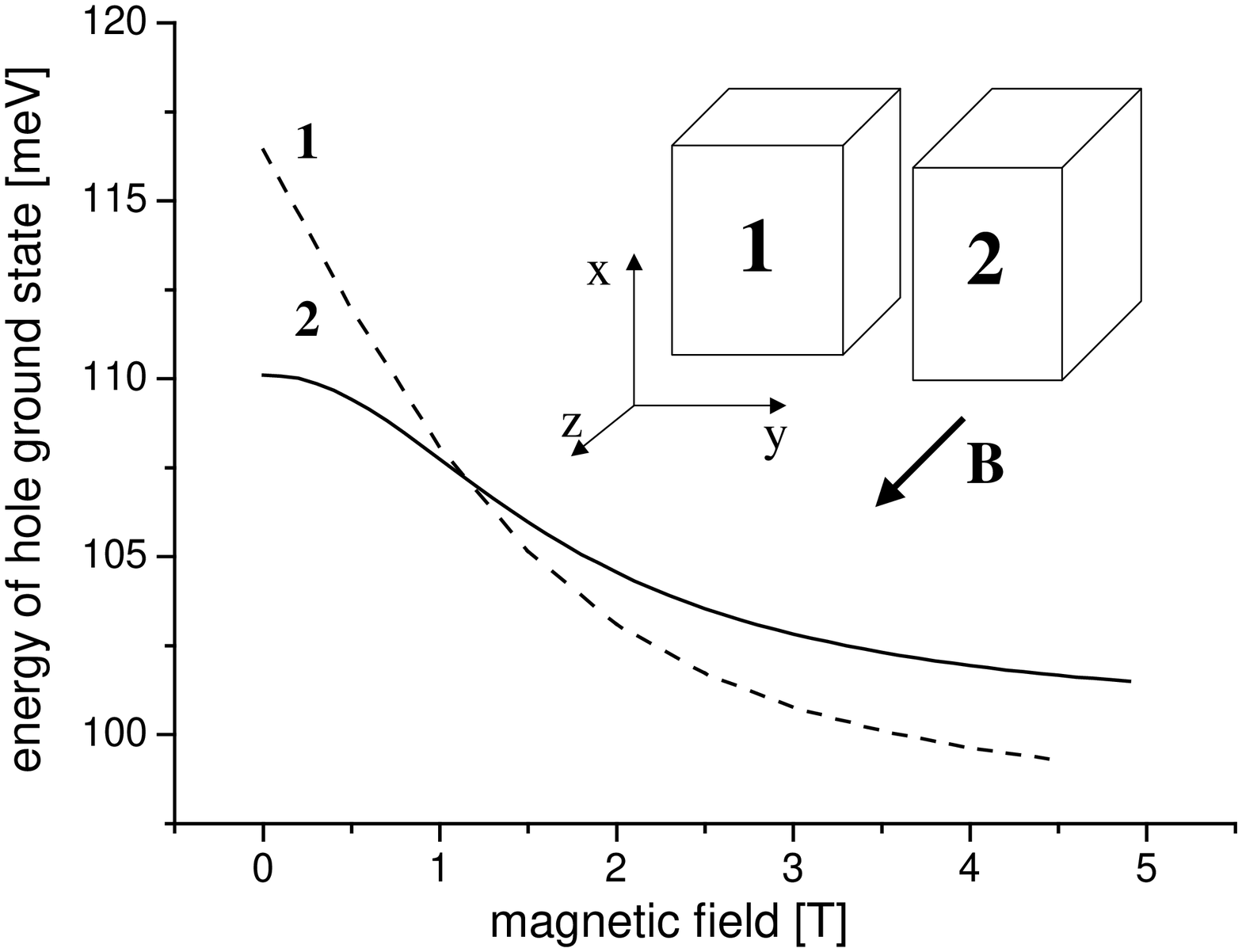}
\caption{\label{transport} Schematic view and the 
evolution of the ground energy of the hole states in
two $\rm Cd_{0.97}Mn_{0.03}Te$ quantum dots in a magnetic
field ${\bf B} \parallel z$. Label 1 corresponds to
$70\AA \times 60\AA \times 50\AA$ quantum dot, label 2 to
$70\AA \times 50\AA \times 65\AA$ quantum dot. The resonant field, that
couples the two quantum dots enabling the hole resonant tunneling,
is about $1.2$ Tesla at T=2K.}
\end{figure}

This sensitivity of the effective g-factor to the exact shape and
orientation of the confining potential can be used to control
hole tunneling between DMS quantum dots.
To clarify this point, let us consider two QDs localizing spatially 
the holes in DMS QW in two in-plane directions. In the
third direction, say $x$-direction, the holes are localized due 
to the presence of a QW confining potential. Such a pair of the QDs
is, thus, made of the same material and have the same
dimensions in the $x$-direction. The remaining two spatial
dimensions of these QDs are chosen in such a way as to result in 
{\em different} expectation values of some component of the angular 
momentum of a hole, say $\langle J_z \rangle$, in the plane of 
the QW. Moreover, the ground
state energy of the hole in the QD having the greater value 
$\langle J_z \rangle$ should be somewhat higher than that
in the other QD, see Fig.~\ref{transport}.
The energy difference between these two ground states and the 
distance between the QDs may prevent the tunneling 
of a hole between these two QDs.
Therefore, we refer to these QDs as uncoupled in this case.
Now, the magnetic field ${\bf B}\parallel z$ shifts the ground states
of holes in QDs. This shift is different for each dot and is
proportional to $\langle \hat{J}_z \rangle$. This is due to 
the fact that the $\langle \hat{J}_z \rangle$ are different 
for our two dots. At some magnetic field energies of ground 
states of holes in two QDs may become the same,
see Fig.~\ref{transport}. One can achieve, thus, the condition 
favorable for the resonant tunneling. 
Let us assume now that the hole originally 
is localized in the first QD (e.g., where the ground state in the
absence of the field is lower) and that the time length
of the magnetic field pulse is made equal to one half 
(three halves, five halves
etc.) of the period of the hole oscillations between the two QDs. 
Then, after such a pulse the hole will be localized in the second QD.

The system is scalable to some extent. It is
possible, for example, to add a third QD with another value 
and/or direction of the resonant magnetic field
that couples the second and the third QD by the tunneling.
As a result, by applying the
magnetic field that effectively couples the second and the third QD
one can also transfer the hole between these dots 
without affecting the hole present in the first QD. Such an 
arrangement can be envisaged as useful in quantum logic gates.

Of course, the quantitative analysis of a possibility of such hole 
tunnel transport as well
as the quantitative description of the Zeeman splitting 
of excitons in DMS QDs requires the consideration of the 
finite potential barriers and
the valence band-conduction band mixing.

\begin{figure}[h]
\epsfysize=6cm \epsfbox{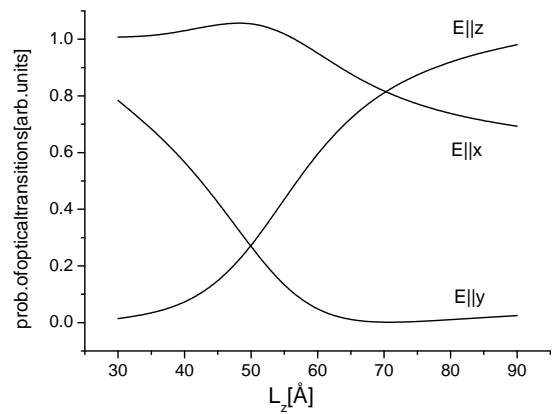}
\caption{\label{polariz_L} The probability of optical
transitions associated with the fourfold degenerate exciton ground state
as a function of $L_z$ for the QD with $L_x=70\AA$ and $L_y=50\AA$.
The three lines correspond to three different linear polarizations 
of the light.}
\end{figure}

Along with the Zeeman splitting, the polarization characteristics
of optical transitions in QDs have to be strongly affected by the
shape of the confining potential. One can expect this even by inspecting
Fig.~\ref{Jz}. A strong dependence of the character of the hole
ground state (i.e., the relative contribution of different spin
components to the wave function) on the shape of a QD should result
not only in a variation of $\langle \hat{J}_z \rangle$ but also
of probabilities of optical transitions.
The values describing the probabilities of various optical
transitions in different linear polarizations of light
associated with the fourfold degenerate exciton ground state
as a function of $L_z$ for QD having $L_x=70\AA$ and $L_y=50\AA$
is plotted in Fig.~\ref{polariz_L}.
With an increase of $L_z$, the situation changes from QW-like
($L_z\ll L_x,L_y$) to QWR-like ($L_z\gg L_x,L_y$). At the same time,
the intensity of the optical transition in the $E\parallel z$
polarization increases, while the intensities of the remaining
two polarizations, considered here, decrease.

\begin{figure}[h]
\epsfysize=6cm \epsfbox{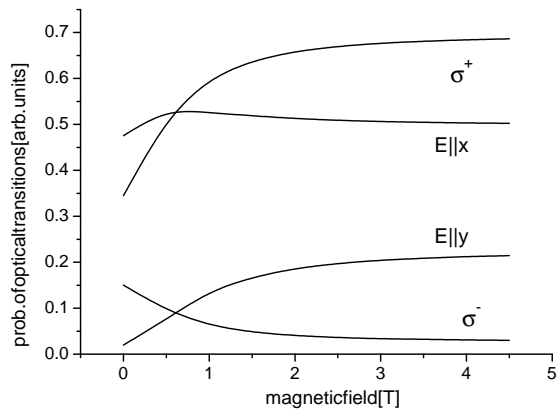}
\caption{\label{polariz_B} The probability of optical transitions in
two circular, and two linear polarizations for the 
ground state of exciton in $\rm Cd_{0.97}Mn_{0.03}Te$ DMS QD
with $L_x=70\AA$, $L_y=60\AA$ and $L_z=50\AA$ in the
magnetic field ${\bf B} \parallel z$ at T=2K.}
\end{figure}

Finally, let us mention yet one more effect that takes place in DMS
quantum dots. It was shown in Ref.\cite{PRB} that the localization
of the holes in two spatial directions in the case of DMS quantum wires
results in a possibility to control the character of the ground state
in the valence band by means of a magnetic field. This leads, in turn,
to a strong magnetic field dependence of the polarization of
optical transitions in such structures. A similar situation occurs,
of course,
in the case of DMS quantum dots. In Fig.~\ref{polariz_B} we plot the
magnetic field dependence of the probabilities of optical transitions in
circular $\sigma^+$, $\sigma^-$, and in two linear polarizations for
ground (non-degenerate) exciton state in $\rm Cd_{0.97}Mn_{0.03}Te$
QD with $L_x=70\AA$, $L_y=60\AA$ and $L_z=50\AA$. It is
seen, that by application of a magnetic field one can effectively control
the polarization characteristics of the light emitted from such DMS QD.

Let us emphasize, that we use in our calculations the full,
anisotropic Luttinger Hamiltonian and a rectangular profile of a QD,
so the selection rules for the optical transitions
differ substantially from those obtained assuming spherically
symmetric quantum dots. For example, in our case the ground 
state of the exciton in a QD is not a dark state. Moreover, all 
valence band states in our case are linear combinations of 
all four different spin (or more appropriately, of all
four projections of the total angular momentum) components, 
resulting in a possibility
of observation of optical transitions in all polarizations.
Nevertheless, in Fig.~\ref{polariz_B} we do not 
plot the probabilities of optical
transitions in $\pi$ polarization ($E\parallel z$). Such transitions
are forbidden, which is connected to the fact, that
we have dealt here with QDs having infinitely high barriers. 
This is analogous to the situation of
QWs, when the optical transitions, such as $e_1$-$h_3$, are
forbidden only in the case of infinite barriers.

The effects, considered in the present paper, do exist also in the
case of non-magnetic quantum dots. However, in the case of diluted
magnetic
semiconductor QD the effects are much more pronounced, as a
direct consequence of the large values of the effective g-factor
of the holes.

\acknowledgments
This work was partially supported by FENIKS RTD 
(EC: G5RD-CT-2001-00535).

\end{document}